\documentclass[]{mn2e}
\usepackage{epsf}
\usepackage[dvips]{epsfig}

\newcommand{\oversim}[2]{\protect{\mbox{\lower0.5ex\vbox{%
   \baselineskip=0pt\lineskip=0.2ex
   \ialign{$\mathsurround=0pt
     #1\hfil##\hfil$\crcr#2\crcr\sim\crcr}}}}}  
\newcommand{\simgreat}{\mbox{$\,\mathrel{\mathpalette\oversim>}\,$}}
\newcommand{\simless}
{\mbox{$\,\mathrel{\mathpalette\oversim<}\,$}} 

\begin{document}
\title[Formation of UCDs]{The Formation of Ultra-Compact 
  Dwarf Galaxies}
\author[M.Fellhauer, P.Kroupa]{Michael Fellhauer and Pavel Kroupa
  \\ 
  Institute for Theoretical Physics and Astrophysics, University
  of Kiel, Germany}
\date{submitted 10.08.2001 -- accepted 26.10.2001} 
\pubyear{2001}

\maketitle

\begin{abstract}
Recent spectroscopic observations of galaxies in the
Fornax-Cluster reveal nearly unresolved `star-like' objects with
red-shifts appropriate to the Fornax-Cluster.  These objects have
intrinsic sizes of $\approx 100$~pc and absolute B-band
magnitudes in the range $-14 < {\rm M}_{\rm B} < -11.5$~mag and
lower limits for the central surface brightness $\mu_{\rm B}
\simgreat 23$~mag/arcsec$^{2}$ (Phillipps et al.\ 2001), and so
appear to constitute a new population of ultra-compact dwarf
galaxies (UCDs). Such compact dwarfs were predicted to form from
the amalgamation of stellar super-clusters (Kroupa 1998), which
are rich aggregates of young massive star clusters (YMCs) that
can form in collisions between gas-rich galaxies.  Here we
present the evolution of super-clusters in a tidal field.  The
YMCs merge on a few super-cluster crossing times.  Super-clusters
that are initially as concentrated and massive as knot~S in the
interacting Antennae galaxies (Whitmore et al. 1999) evolve to
merger objects that are long-lived and show properties comparable
to the newly discovered UCDs.  Less massive super-clusters
resembling knot~430 in the Antennae may evolve to
$\omega$~Cen-type systems.  Low-concentration super-clusters are
disrupted by the tidal field, dispersing their surviving star
clusters while the remaining merger objects rapidly evolve into the
$\mu_{\rm B}-M_{\rm B}$ region populated by low-mass Milky-Way
dSph satellites. 
\end{abstract}

\begin{keywords}
  galaxies: interactions -- galaxies: formation -- galaxies: star 
  clusters -- galaxies: dwarfs -- globular clusters: $\omega$ Cen 
  (NGC~5139) -- methods: N-body simulations
\end{keywords}

\section{Introduction}
\label{sec:intro}
\begin{figure*}
  \begin{center} 
    \epsfxsize=15cm 
    \epsfysize=15cm 
    \epsffile{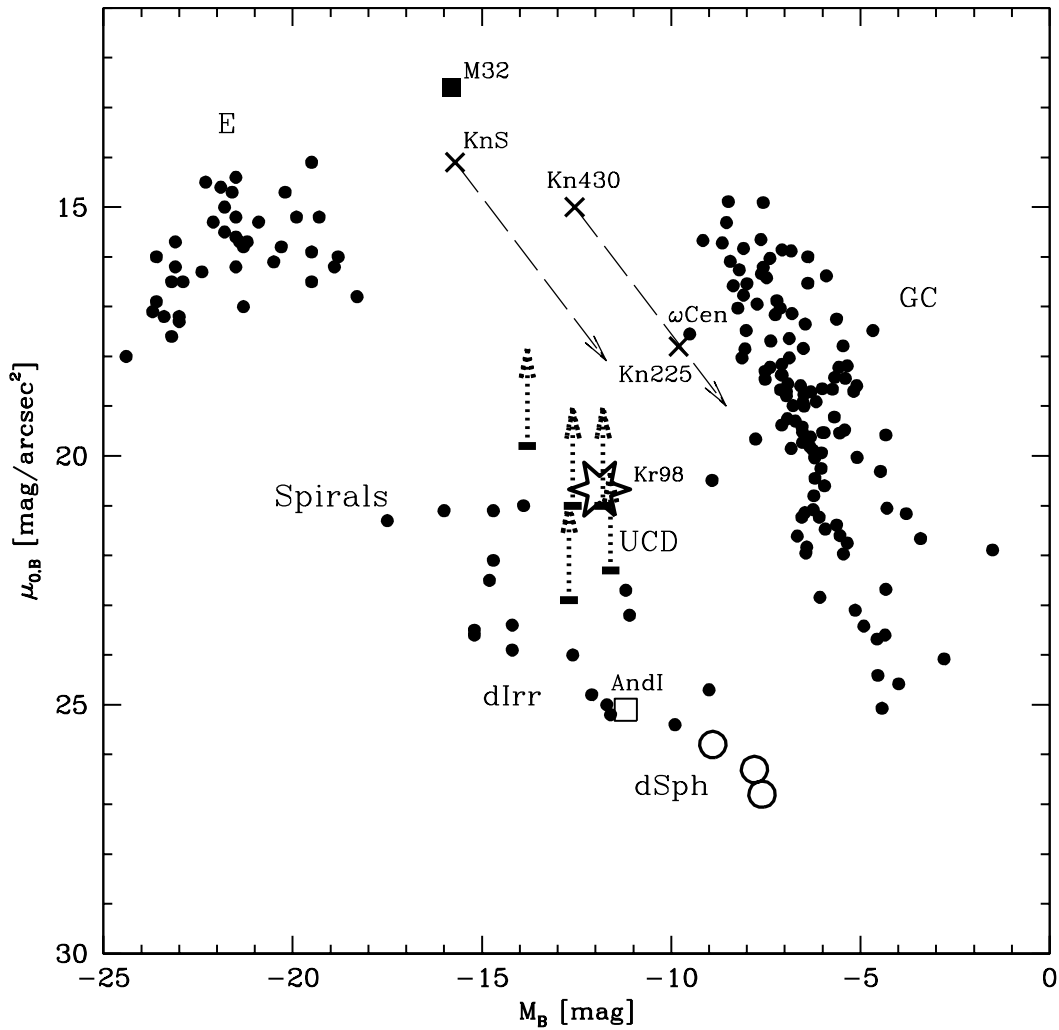}
    \caption{Central surface brightness, $\mu_{0,{\rm B}}$, vs absolute
      photometric B-band magnitude, $M_{\rm B}$ (the Kormendy
      diagram).  The solid circles are data for Milky-Way globular
      clusters (GC, Harris 1996), Local-Group dwarf galaxies (dIrr, AndI
      and dSph, Mateo 1998), elliptical galaxies (E, Peletier et al.\
      1990).  The positions of disk galaxies are indicated by
      ``Spirals'' (cf. Ferguson \& Binggeli 1994).  The newly discovered
      UCDs are shown by arrows with lower-limits on $\mu_{\rm B}$
      (Phillipps et al.\ 2001). Three ``knots'', (KnS, Kn430, Kn225,
      tables~1 and~2 in Whitmore et al. 1999), observed in the
      interacting Antennae galaxies are shown as thick crosses.  Knots
      KnS and Kn430 are roughly 10~Myr old so that $M/L_{\rm
        V}\approx0.076$ (fig.~7 in Smith \& Gallagher 2001), while Kn225
      is about 500~Myr old.  The fading of all three knots to $M/L_{\rm
        B}=3$ due to stellar evolution but assuming no morphological
      evolution (eq.~\ref{eqn:morph}), is indicated by the dashed
      arrows.  The thick six-pointed star shows the merged
      super-clusters predicted by Kroupa (1998) to result from the
      dynamical evolution of such knots, which consist of dozens to
      hundreds of young star clusters.}  
    \label{fig:k_obs} 
  \end{center}
\end{figure*}

In a recent spectroscopic survey of all objects around NGC~1399 in the
Fornax-Cluster, carried out to find all dwarf galaxies, Phillipps et
al.\ (2001) report five ultra-compact dwarf galaxies (UCDs; see also 
Hilker et al.\ 1999).  These
have spectra typical for late-type metal-rich and old stellar
populations, are marginally resolved and have red-shifts comparable to
the Fornax-Cluster, ruling out either faint background galaxies or
foreground stars.  The UCDs have intrinsic sizes of around 100~pc,
absolute B-band magnitudes of $-14 < {\rm M}_{\rm B} < -11.5$~mag and
lower limits (due to failure to resolve their cores) of the central
surface-brightness $\mu_{\rm B} \simgreat 23$~mag/arcsec$^{2}$.
Further analysis of photographic plates show no sign of low-luminosity
envelopes around these objects which rules out the possibility that
they are nucleated dwarf ellipticals (dE,N) with faint envelopes.
Therefore, these objects are either extremely compact dwarf galaxies
or extremely large and massive ($10^7-10^8\,M_\odot$, assuming
$M/L_{\rm B}=3$) star clusters.  In the $\mu_{\rm B}-M_{\rm B}$
diagram they fall in the empty region between `ordinary' dwarf
galaxies and globular clusters (Fig.~\ref{fig:k_obs}). 

High-resolution HST-images of the star forming regions in interacting
galaxies like the Antennae (Whitmore et al.\ 1999; Zhang \& Fall 1999)
or Stephan's Quintet (Gallagher et al.\ 2001) resolve some of these
regions into dozens to hundreds of young massive star clusters.
According to Whitmore et al.\ the individual clusters have effective
radii of about 4~pc and masses $10^{4}$--$10^{6}$~M$_{\odot}$, with a
mass-spectrum following a steep power law $\Psi_{M} \propto M^{-2}$
(Zhang \& Fall 1999).  The striking point is that these young star
clusters are themselves clustered into groups of a few to several
hundred star clusters spanning projected regions of a few~100~pc, with
a higher cluster concentration at the centre.  Measurements of the
relative velocities between the star clusters within such objects are
now becoming available.  Preliminary results indicate $\approx20$~km/s
(B.~Whitmore, private communication), which is consistent with virial
masses $\approx10^7\,M_\odot$.  Age determinations in the Antennae
show that these star clusters are extremely young (3--7~Myr).  While
in the Antennae young massive star clusters are preferably found in
the central region of the interacting pair, NGC~7319 in Stephan's
Quintet has young star clusters in the long tidal arm and the
intra-group region north of NGC~7319 (Gallagher et al.\ 2001). 

Examples of clusters of star clusters are shown in
Fig.~\ref{fig:k_obs} as thick crosses.  Of these, knot ``S'' is
particularly remarkable, as its mass is estimated to be about
$2.1\times10^7\,M_\odot$ assuming it has a mass-to-light ratio in
the photometric V-band of $M/L_{\rm V}=0.076$ for an age of
10~Myr (fig.~7 in Smith \& Gallagher 2001) and $M_{\rm V}=-15.8$
(section~4.4.2 in Whitmore et al. 1999).  Two other knots (Kn430
and Kn225) are plotted in Fig.~\ref{fig:k_obs} with fading
vectors to $M/L_{\rm B}=3$ to indicate their position after a few
Gyr if the shape of their density profile remains unchanged,
\begin{eqnarray}
  {L_{\rm c} \over L_{\rm tot}} & = & {\rm constant},
  \label{eqn:morph}
\end{eqnarray} 
where $L_{\rm c}$ and $L_{\rm tot}$ are the central and
integrated luminosity, respectively. 

The dynamical evolution of these knots is an interesting problem
to consider because the merging of many star clusters is not well
understood but is a pre-requisite for the understanding of the
future fate of such knots.  By scaling the numerical results of
Garijo, Athanassoula \& Garcia-Gomez (1997, hereinafter GAG), who
studied the formation of cD galaxies from clusters of 50
equal-mass galaxies, to the present problem of a super-cluster
containing 50 star clusters, each cluster having a mass $M_{\rm
  cl}=10^6\,M_\odot$, Kroupa (1998) studied the dynamical
evolution of such ``stellar super-clusters''.  These are not to
be confused with the commonly-used terminology
`super-star-cluster', SSC, which is one young star cluster with a
mass typical for globular clusters (e.g. Smith \& Gallagher
2001).  Depending on the initial concentration of such an object,
the majority of star-clusters merge within a few super-cluster
crossing times.  Within 100~Myr a stellar system forms which has
a relaxation time significantly longer than a Hubble time.  It is
thus a low-mass spheroidal dwarf galaxy (as opposed to the
well-known dwarf-spheroidal satellite galaxies, denoted dSph),
with a half-mass radius of between about~45 to 100~pc, possibly
with a high specific frequency of globular clusters if some
survive the merging phase.  From figs.~16-18 in GAG the central
surface density for essentially all models (collapsing
super-clusters and super-clusters initially in virial
equilibrium) is $1$~mass-unit/length-unit$^2$.  This translates to
$\mu=1000\,M_\odot/$pc$^2$ and to the apparent-magnitude surface
brightness $\mu_{\rm B}=19.48$~mag/arcsec$^2$ using 
\begin{eqnarray}
  \mu_{\rm B} [{\rm mag/arcsec}^2] & = & - \ 2.5\,{\rm log}_{10}
  \left( \mu \, \frac{L_{\rm B}} {M}
  \right) \\
  & & + \ M_{{\rm B},\odot} \ - \ 5 \ + \ {\cal S}, \nonumber
  \label{eqn:muB}
\end{eqnarray}
where $M_{{\rm B},\odot}=+5.41$ is the absolute B-band magnitude
of the Sun, and ${\cal S}=26.572=2.5$log$_{10}f^2$ with
$f=2.06\times10^5$ arcsec per radian, and assuming the B-band
mass-to-light ratio $M/L_{\rm B}=3$ in solar units.  Assuming
that 50--100~per cent of all clusters merge to form the
spheroidal dwarf galaxy, its integrated absolute B-band
magnitude, 
\begin{eqnarray}
  M_{\rm B} & = & - \ 2.5\,{\rm log}_{10} \left( M  \,
    \frac{L_{\rm B}} {M} \right) + M_{{\rm B},\odot},
\label{eqn:Mb}
\end{eqnarray}
where $M$ is the mass of the object in $M_\odot$, lies in the
range $-12.64\le M_{\rm B}\le -11.89$.

Fig.~\ref{fig:k_obs} shows that the predicted spheroidal dwarf
galaxies fall in the region of the Kormendy diagram that the UCD data
of Phillipps et al.\ (2001) cover, nicely verifying the prediction.
The hypothesis that the UCDs were formed as super-clusters appears to
be a reasonable possibility, since gas-rich galaxies will have merged
in profusion in sub-groups during the early phases of the formation of
the Fornax galaxy cluster.  The UCDs may thus have formed during the
coalescence of a subgroup of late-type gas-rich galaxies, which was
infalling into the Fornax cluster. Tidal shocking of such UCDs and/or
initially less-massive super-clusters may account for some if not the
majority of known spheroidal dwarf galaxies found in large numbers in
rich galaxy clusters (Gallagher, Conselice \& Wyse 2001).

Kroupa (1998) studied isolated super-clusters, but tidal shaping
of the merger objects may be significant.  In this contribution we
report self-consistent stellar-dynamical calculations of the
merging process in super-clusters whilst they are immersed in the
tidal field of a parent galaxy.  A detailed study of the merging
time-scales and efficiencies for super-clusters on circular
orbits shows that the star clusters merge very efficiently on a
few dynamical time-scales of the super-cluster (Fellhauer et al.\
2001).  In this paper we focus on a comparison with the recently
discovered UCDs and perform computations of super-clusters with
varying initial concentrations on eccentric orbits. 

The models are explained in Sect.~\ref{sec:setup} where a brief
description of the N-Body code is given.  In Sect.~\ref{sec:results}
the properties of the merger objects on circular orbits are briefly
described and a model on an eccentric orbit is discussed in
detail. The evolution of a low-mass extended merger object is also
studied.  Conclusions are presented in Section~\ref{sec:conclus}.


\section{Numerical Method and Setup}
\label{sec:setup}


\subsection{The code}
\label{sec:code}

The orbital integration of the particles is performed with the
particle-mesh code {\sc Superbox} (Fellhauer et al.\ 2000).  In {\sc
Superbox} densities are derived on Cartesian grids using the
nearest-grid-point (NGP) scheme.  From these density arrays the
potential is calculated via the fast Fourier-transformation.  Forces
are obtained using higher-order differentiation based on the NGP
scheme but comparable in precision with standard CIC (cloud-in-cell)
algorithms.  The particles are integrated using a fixed time-step
Leap-Frog algorithm.  For an improved resolution at the regions of
interest, a hierarchical grid-architecture with two levels of high
resolution sub-grids are used for each star cluster in the
super-cluster.  These sub-grids track the density maxima of the
individual star clusters, and are adjusted at the beginning of the
computation to meet individual requirements. 

The sizes of the two high resolution grid-levels are chosen such that
the innermost grids with the highest resolution (1~pc per grid-cell
length) cover the star clusters, while the medium-resolution grids
cover the merger object to ensure that the forces between each cluster
are treated at least at medium resolution.  Finally, the outermost
grid, which stays fixed and has the lowest resolution, covers the
whole orbit of the super-cluster around the parent galaxy.  The
time-step used in the calculations is $dt = 0.1$~Myr, which is
significantly shorter than the crossing time in the star-clusters
($t_{\rm cr} \approx 6$~Myr).

Out notation is as follows: we use large letters for galactocentric
coordinates ($X, Y, Z, R$) and derived quantities, and small letters
($x, y, z, r$) for super-cluster-centric coordinates and derived
quantities.


\begin{table*}
  \begin{minipage}[h!]{15cm}
  \begin{center}
    \caption{Properties and orbits of the model super-clusters.
      The columns give the following data: name of the model,
      number of star clusters in the super-cluster,
      Plummer-radius of the super-cluster, crossing time of the
      super-cluster, total mass of the super-cluster,
      perigalacticon distance, apo-galacticon distance, orbital
      period, concentration parameter, strength of tidal field
      parameter.}
    \vspace*{0.5cm}
    \label{tab:mods}
    \begin{tabular}[h!]{lrrrrrrrrr} \hline
      name & $N_{0}$ & $r_{\rm pl}^{\rm sc}$ & $t_{\rm cr}^{\rm
        sc}$ & $M_{\rm sc}$ & $R_{\rm peri}$ & $R_{\rm apo}$ &
      $T_{\rm orb}$ & $\alpha$ & $\beta_{\rm peri}$ \\
       & & [pc] & [Myr] & [$10^{7}$M$_{\odot}$] & [kpc] & [kpc]
       & [Myr] & & \\ \hline
      SD05A & 20 & 50 & 7.4 & 2.0 & 10 & 10 & 300.7 & 0.12 & 0.62
      \\
      SD06A & 20 & 75 & 13.5 & 2.0 & 10 & 10 & 300.7 & 0.08 &
      0.93 \\
      SD09A & 20 & 50 & 7.4 & 2.0 & 20 & 20 & 569.6 & 0.12 & 0.42
      \\
      SD10A & 20 & 75 & 13.5 & 2.0 & 20 & 20 & 569.6 & 0.08 &
      0.60 \\
      SD11 & 20 & 150 & 38.3 & 2.0 & 20 & 20 & 569.6 & 0.04 &
      1.28 \\
      SD17B & 20 & 50 & 7.4 & 2.0 & 50 & 50 & 1400.7 & 0.12 &
      0.23 \\
      SD18A & 20 & 75 & 13.5 & 2.0 & 50 & 50 & 1400.7 & 0.08 &
      0.35 \\
      SD19E & 20 & 150 & 38.3 & 2.0 & 50 & 50 & 1400.7 & 0.04 &
      0.70 \\ \hline
      R06 & 20 & 300 & 108.4 & 2.0 & 30 & 60 & 900 & 0.02 & 2.0
      \\ \hline
      E01 & 32 & 50 & 9.4 & 1.2 & 2 & 20 & 250 & 0.08 & 1.29 \\
      \hline
    \end{tabular}
  \end{center}
  \end{minipage}
\end{table*}

\subsection{Orbits}
\label{sec:orbit}

The super-cluster orbits within an external potential of a
parent galaxy given by
\begin{eqnarray}
  \label{eq:galpot}
  \Phi(R) & = & \frac{1}{2} V_{0}^{2} \cdot {\rm ln}(R_{\rm
    gal}^{2} + R^{2})
\end{eqnarray}
with core radius $R_{\rm gal} = 4$~kpc and circular velocity $V_{0} =
220$~km/s.  Dynamical friction is also analytically added using
the Chandrasekhar expression (Binney \& Tremaine, their eq.~7-18):
\begin{eqnarray}
  \label{eq:dynfric}
  \frac{{\rm d}\vec{V}_{M}} {{\rm d}t} & = & - \ \frac{4 \pi \ln(
    \Lambda) G^{2} \varrho_{\rm gal} M_{\rm sc}} {V_{M}^{3}}
  \times \\ 
  & & \left[ {\rm erf} \left( \frac{V_{m}} {\sqrt{2} \sigma}
    \right) \ - \ \frac{ \sqrt{2} V_{M}} {\sqrt{\pi} \sigma} \exp
    \left(-\frac{V_{M}} {\sqrt{2} \sigma} \right)  \right] \cdot
    \vec{V}_{M} \nonumber
\end{eqnarray} 
with $\ln(\Lambda)=1.5$ (Velazquez \& White 1999; Fellhauer et
al.\ 2000) the so-called Coulomb-logarithm, $\varrho_{\rm gal}$ is  
the density of the galactic halo at the position of the super-cluster, 
$M_{\rm sc}$ is the total mass of the super-cluster, $V_{M}$ is the 
orbital velocity of the super-cluster in the tidal field of the host 
galaxy and $\sigma$ is the velocity dispersion in the galactic halo
at the position of the super-cluster. 

We adopt the tidal field of a Milky-Way type galaxy, but our results
are applicable to a galaxy cluster such as Fornax in the sense that
super-clusters that remain stable over many highly eccentric orbits
will also survive in the potential well of Fornax, whereas
super-clusters that are torn apart immediately are not likely to
survive in a galaxy cluster.  Gallagher, Conselice \& Wyse (2001)
discuss dwarf-galaxy survival in more detail, and we note that
spheroidal dwarfs without dark matter can readily survive passages
through the centres of galaxy-clusters owing to the large encounter
velocities.  We do not aim to model the UCDs observed in the Fornax
cluster in detail, which is impossible in any case given that little
is known about their orbits, but merely concentrate on the generic
properties of the merger objects when they are subject tidal fields
that vary in strength.

To study the merging time-scales and efficiencies circular orbits
are adopted where the centre of the super-cluster is placed at a
distance $R=D$ from the galactic centre.  The distance $D$ is
varied in the range of 5--100~kpc.  A detailed discussion of the
involved time-scales can be found in Fellhauer et al. (2001).  To
study the evolution of the merger objects in a time-dependent
tidal field the parameter space is extended to eccentric orbits.


\subsection{Super-clusters}
\label{sec:sc}

A range of models (Table~\ref{tab:mods}) are setup to investigate
the importance of the tidal field and initial concentration of
the super-cluster. 

The young massive star clusters are modelled as Plummer-spheres
with Plummer-radii $r_{\rm pl}=$ 4--6~pc and cutoff radii
20--30~pc.  The clusters all have masses of
$10^{6}$~M$_{\odot}$ if no mass spectrum is taken into account, or
$10^{6}$~M$_{\odot}$ and $3.2 \cdot 10^{5}$~M$_{\odot}$ to model a 
mass spectrum.  Estimating the relaxation time $t_{\rm relax}$ of 
these star clusters (Spitzer \& Hart 1971),
\begin{eqnarray}
  \label{eq:relax}
  t_{\rm relax} & = & 0.138 \cdot \frac{\sqrt{M_{\rm cl}}
    r_{h}^{3/2}} {\left< m \right> \sqrt{G} \ln(0.4\,n)}.
\end{eqnarray}
where $M_{\rm cl}$ is the mass of the star cluster, $r_{h}$ is the 
half-mass radius, the average stellar mass $\left< m \right> = 
0.4$~M$_{\odot}$ (using the universal IMF, Kroupa 2001) and $n$ 
the number of stars.  This leads to $t_{\rm relax}\approx$4~Gyr for 
the more massive star clusters and 2.5~Gyr for the less massive 
ones.  These time-scales are
much longer than the cluster--cluster merging time-scale.  The
resulting merger objects have relaxation times longer than a Hubble
time.  Surviving star clusters are only counted and their position is
traced; their internal evolution is not studied here.  Therefore it is
possible to carry out the orbital integration with a particle-mesh
code which suppresses two-body relaxation processes. 

To model a super-cluster, $N_{0}$, of these clusters are placed in a
Plummer distribution with Plummer radius ($r_{\rm pl}^{\rm sc}$)
ranging from 50--300~pc and cut-off radius ($r_{\rm cut}^{\rm sc}$).
The super-cluster is initially in virial equilibrium.

To describe our parameters with dimensionless variables we define
\begin{eqnarray}
  \label{eq:alpha}
  \alpha & = & r_{\rm pl} / r_{\rm pl}^{\rm sc}, \\
  \label{eq:beta}
  \beta & = & r_{\rm cut}^{\rm sc} / r_{\rm tidal},
\end{eqnarray}
where $r_{\rm tidal}$ is the tidal radius of the super-cluster.
In the case of eccentric orbits the tidal radius is estimated at
perigalacticon, 
\begin{eqnarray}
  \label{eq:tidal}
  r_{\rm tidal} & = & \left (\frac{M_{\rm sc}}{3\,M_{\rm gal}(R)}
  \right)^{1/3} \, R_{\rm peri}.
\end{eqnarray}
While $\alpha$ describes how densely the super-cluster is filled
with star clusters, $\beta$ indicates the strength of the tidal
forces acting on the super-cluster.  With our choices of the
parameters $r_{\rm pl}^{\rm sc}$ and $D$, $\alpha$ ranges from
0.02 to 0.12 and $\beta$ falls in the interval~0.2 to~2.7.

\begin{table*}
  \begin{minipage}[h!]{15cm}
    \begin{center}
      \caption{Sizes of the merger objects, measured at $t=1$~Gyr
        and at $t=5$~Gyr for different choices of $\alpha$ and
        $\beta$.  Only the bound mass is taken into account 
        (cf. Fig.~\ref{fig:lag}).}
      \vspace*{0.5cm}
      \label{tab:size}
      \begin{tabular}[h!]{lllrrrrrr} \hline
        name & $\alpha$ & $\beta$ & \multicolumn{2}{c}{Half-Mass
          Radius [pc]} & \multicolumn{2}{c}{90~\%-Mass Radius
          [pc]} & \multicolumn{2}{c}{Tidal Radius [pc]} \\ 
         & & & $t=1.0$~Gyr & $t=5.0$~Gyr & $t=1.0$~Gyr &
         $t=5.0$~Gyr & $t=1.0$~Gyr & $t=5.0$~Gyr \\ \hline
        SD05A & 0.12 & 0.62 &  39.0 &  40.5 &   123.2 &   135.4 &
        400 &     400 \\ 
        SD06A & 0.12 & 0.42 &  48.0 &  50.2 &   151.5 &   150.9 &
        600 &     600 \\ 
        SD09A & 0.12 & 0.23 &  38.3 &  39.0 &   125.5 &   124.2 &
        1,120 & 1,120 \\  
        SD10A & 0.08 & 0.93 &  79.2 &  77.3 &   200.0 &   188.9 &
        390 &     390 \\ 
        SD11  & 0.08 & 0.60 &  43.0 &  44.5 &   146.1 &   150.9 &
        610 &     610 \\ 
        SD17B & 0.08 & 0.35 &  71.2 &  72.4 &   230.4 &   221.5 &
        1,110 & 1,110 \\  
        SD18A & 0.04 & 1.28 & 181.2 & 144.6 &   491.1 &   348.1 &
        580 & 580 \\ 
        SD19E & 0.04 & 0.70 & 236.6 & 163.2 & 1,058.8 &   576.5 &
        1,010 &   910 \\ \hline
        R06   & 0.02 & 2.00 & --- & --- & --- & --- & --- & ---
        \\ \hline
        E01   & 0.08 & 1.29 & 54.6 & 41.9 & 149.2 & 146.3 & 188 &
        152 \\ \hline
      \end{tabular}
    \end{center}
  \end{minipage}
\end{table*}

The relaxation time of the cluster system is very short, i.e. using
Eq.~\ref{eq:relax} with $n=N_{0}$ would lead to a nominally shorter 
relaxation time  
than the crossing time of the super-clusters (about 0.3--0.4 
$t_{\rm cr}^{\rm sc}$).  This shows that such systems are heavily
collision-dominated and therefore the merging process is violent and 
the merging timescales should be very short. 


\section{Results}
\label{sec:results}

As discussed by Kroupa (1998) and Fellhauer et al. (2001) the
evolution of the young super-cluster is dominated initially,
during the violent merging time, by cluster--cluster merging and
interaction processes.  The merging efficiency is, in general,
very high.  As long as the cutoff radius is smaller than the
tidal radius ($\beta \leq 1.0$) almost all star clusters merge.
The total number of clusters that merge decreases linearly and
slowly with increasing $\beta>1$. 


\subsection{Circular orbits}
\label{sec:circ}

The merger objects consisting initially of $N_{0}=20$ massive star
clusters (each with $10^{6}$~M$_{\odot}$) on circular orbits have
half-mass radii and sizes summarised in Table~\ref{tab:size}.  The
time-scale for the merging process is very short.  Within $5-10\times
t_{\rm cr}^{\rm sc}$ ($t_{\rm cr}^{\rm sc}$ being the internal
crossing time of the super-cluster) the initial violent merging phase
is over.  During this phase the number of remaining clusters decreases
exponentially, the decay thus being similar to the radioactive decay
law.  For initially concentrated super-clusters ($\alpha \geq 0.08$)
this phase lasts $\simless100$~Myr.  The resulting merger object is
very dense and stable over 10~Gyr. In Fig.~\ref{fig:k_obs} it
coincides with the Kr98 prediction. 

As long as $0.08\simless \alpha \simless 0.12$ there is almost no
evolution in mass and size of the merger object.  If $\alpha
\simless 0.04$ or $\beta > 1.0$ the merger object looses a part
of its mass and shrinks until it reaches a stable configuration,
and is associated with a population of surviving star clusters
that spread out along the orbit.  The merger object lies close to
the Kr98 prediction in Fig.~\ref{fig:k_obs} despite the mass
lost.

Between 70 and 95~\% of the mass of the super-cluster ends up in
the merger objects, which have central densities of about
650~M$_{\odot}$/pc$^{3}$ and exponential scale lengths of about
9~pc.  The surface density can be well fitted by a
de-Vaucouleur-profile, the line-of-sight velocity dispersion is
about 20~km/s.  The shape of the objects is spherical with
axes-ratios larger than 0.9.  


\subsection{Dense super-cluster on an eccentric orbit}
\label{sec:ecchd}

\begin{figure}
  \begin{center} 
    \epsfxsize=08.0cm 
    \epsfysize=08.0cm
    \epsffile{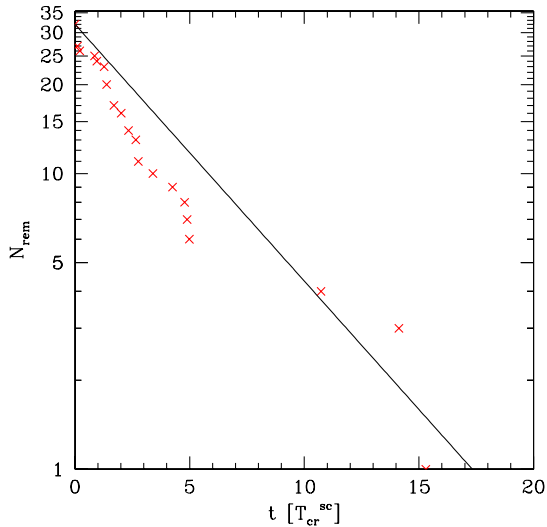} 
    \caption{The number of remaining star clusters, $N_{\rm rem}$, vs.\ 
      time for model E01.  Within the first 5 crossing times of the 
      super-cluster the merging can be fitted with an exponential with 
      an exponential decay time of about 47~Myr.  This is the same fit
      as already obtained by Fellhauer et al.\ (2001) for a large variety 
      of super-clusters. }  
    \label{fig:merger} 
  \end{center}
\end{figure}

The models are extended to include eccentric orbits to study the
influence of tidal heating acting on the super-clusters and the
resulting merger objects (Table~\ref{tab:mods}).  As reference model
in this paper we start with a dense super-cluster containing $N_0=32$
star clusters, 3 having a mass of $10^{6}$~M$_{\odot}$ and 29 having
$3.2 \cdot 10^{5}$~M$_{\odot}$ to mimic a mass spectrum of the form
$\Psi_{M} \propto M^{-2}$.  The star clusters have a Plummer radius of
$r_{\rm pl} = 4$~pc and crossing times of 0.75 and 1.32 Myr,
respectively.  The super-cluster they are distributed in, has a total
mass $M_{\rm sc} = 1.23 \times 10^{7}$~M$_{\odot}$, a Plummer radius
$r_{\rm pl}^{\rm sc} = 50$~pc, a cut-off radius of $r_{\rm cut}^{\rm
sc} = 250$~pc (giving $\alpha = 0.08$) and a crossing time of 9.4~Myr.
The super-cluster is placed at apo-galacticon ($D = 20$~kpc) on a
highly eccentric orbit with eccentricity $e=0.82$, where
\begin{eqnarray}
  e & = & \frac{R_{\rm apo}-R_{\rm peri}}{R_{\rm apo}+R_{\rm
      peri}}.  
\label{eqn:ecc}
\end{eqnarray}
At perigalacticon $\beta_{\rm peri} = \beta = 1.29$.  However,
since the orbit starts at apo-galacticon and has an orbital
period $T_{\rm orb} \approx 250$~Myr the super-cluster has
$\beta<1$ for a time-span $ n_t \times t_{\rm cr}^{\rm sc}$, $n_t
\approx 10 $, giving enough time for the super-cluster to evolve
essentially as the models studied above.  Most star clusters
merge within $5 \times t_{\rm cr}^{\rm sc}$ (Fig.~\ref{fig:merger}). The
number of star clusters decreases exponentially,
\begin{eqnarray}
  \label{eq:merger}
  N(t) & = & N_{0} \cdot \exp \left( - \frac{t}{\tau} \right),
\end{eqnarray}
with $t$ measured in units of the crossing time of the super-cluster
and $\tau = 5\,t_{\rm cr}^{\rm sc}$ which corresponds to 47~Myr for model 
E01.  This exponential decrease is shown as a straight line in
Fig.~\ref{fig:merger}.  The initial violent merging phase is dominated
by clusters on highly eccentric orbits within the super-cluster. Star
clusters initially on near-circular orbits in the super-cluster
survive until $\simgreat 5\,t_{\rm cr}^{\rm sc}$, but ultimately they
merge due to dynamical friction within the super-cluster.

\begin{figure}
  \begin{center} 
    \epsfxsize=08.0cm 
    \epsfysize=08.0cm
    \epsffile{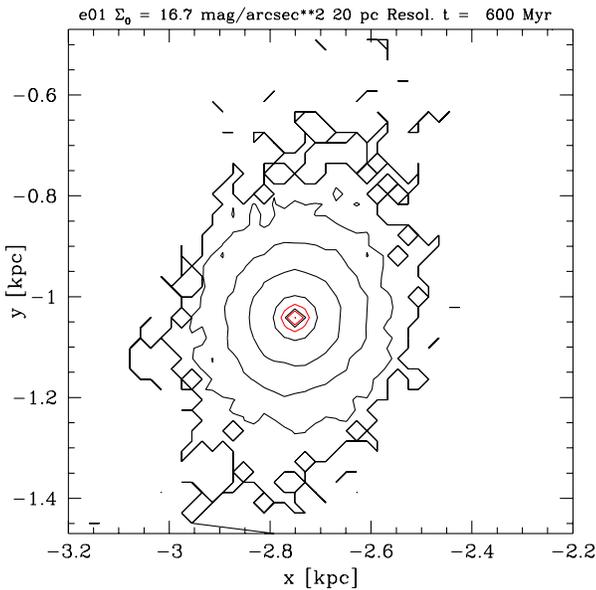} 
    \caption{Contour plot of the merger object E01 at 600~Myr
      shortly after the violent merging phase is over.  While the
      central area is circular, the envelope shows an ellipticity
      of about 0.6.}  
    \label{fig:c600} 
  \end{center}
\end{figure}

\begin{figure}
  \begin{center}
    \epsfxsize=08.0cm
    \epsfysize=08.0cm
    \epsffile{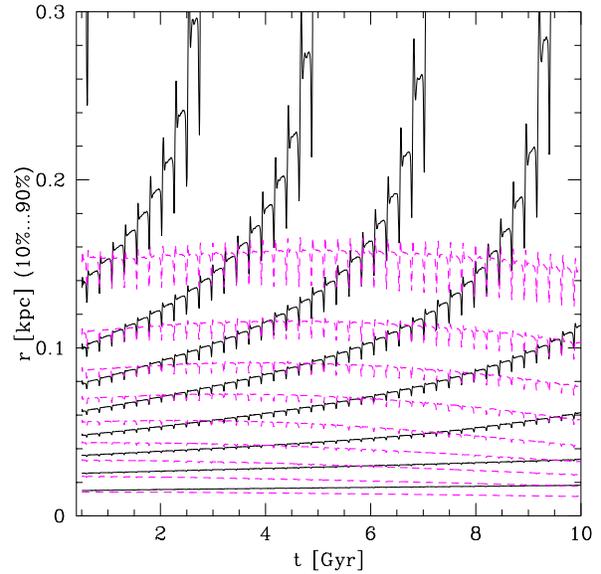}
    \caption{Lagrangian radii containing 10, 20, ..., 90 \% of the mass
      of merger object E01 vs.\ time.  Solid lines are all
      particles, dashed lines are the bound particles only.} 
    \label{fig:lag}
  \end{center}
\end{figure}

\begin{figure}
  \begin{center}
    \epsfxsize=08.0cm
    \epsfysize=08.0cm
    \epsffile{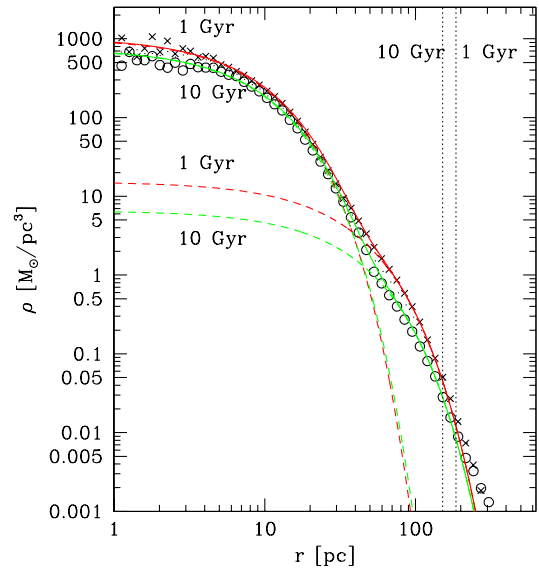}
    \caption{Radial density profile of merger object E01
      at $t=$ 1
      (crosses), 5 (points) and 10 Gyr (circles).  Profiles can
      be fitted by two exponentials (dashed lines, according to
      eq.~\ref{eq:rho}) with the overall fitting function as
      solid line.  Vertical lines show the tidal radii at $t=$ 1
      and 10~Gyr derived from Eq.~\ref{eq:tidal}.}
    \label{fig:dens}
  \end{center}
\end{figure}

\begin{figure}
  \begin{center}
    \epsfxsize=08.0cm
    \epsfysize=08.0cm
    \epsffile{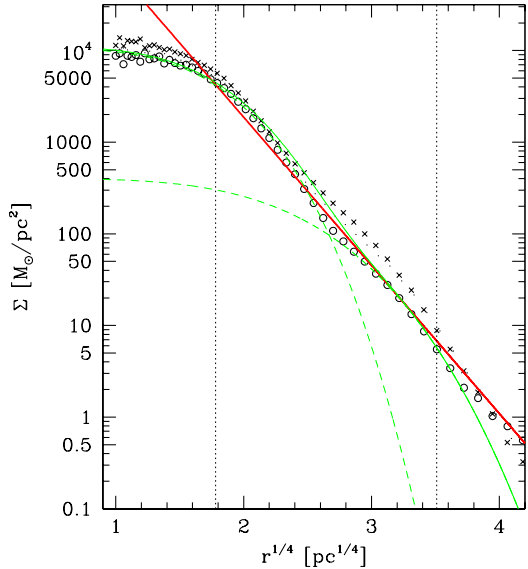}
    \caption{Surface density profile of the merger object E01.
      Profiles are displayed at $t =$ 1 (crosses), 5 (dots), 10
      Gyr (circles). Solid thick line shows a de-Vaucouleurs
      fit to the data at $t=10$~Gyr.  Also shown as dashed lines
      are two exponentials as a fitting function with the sum of
      both shown as thin solid curve. Right vertical dotted line
      shows the tidal radius taken from Eq.~\ref{eq:tidal}, left
      vertical dotted line shows the core radius (both at
      $t=$~10~Gyr).}  
    \label{fig:surf}
  \end{center}
\end{figure}

\begin{figure}
  \begin{center}
    \epsfxsize=08.0cm
    \epsfysize=08.0cm
    \epsffile{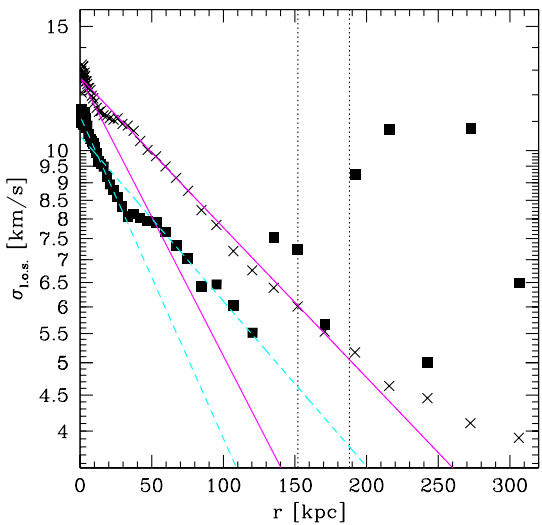}
    \caption{Line-of-sight velocity dispersion profiles of 
      merger object E01 at $t =$ 1 (crosses) and 10 Gyr
      (boxes). Profiles can be fitted by exponentials.  Fitting
      parameters for $t=1$~Gyr: inner: $\sigma_{0} = 12.7 \pm
      0.1$~km/s, $r_{\rm exp} = 110 \pm 10$~pc; outer:
      $\sigma_{0} = 12.4 \pm 0.2$~km/s, $r_{\rm exp} = 222 \pm
      5$~pc; $t=10$~Gyr: inner $\sigma_{0} = 11.6 \pm 0.6$~km/s,
      $r_{\rm exp} = 90 \pm 2$~pc; outer: $\sigma_{0} = 10.5 \pm
      0.3$~km/s, $r_{\rm exp} = 185 \pm 10$~pc.}  
    \label{fig:losv}
  \end{center}
\end{figure}

\begin{figure}
  \begin{center}
    \epsfxsize=08.0cm
    \epsfysize=09.5cm
    \epsffile{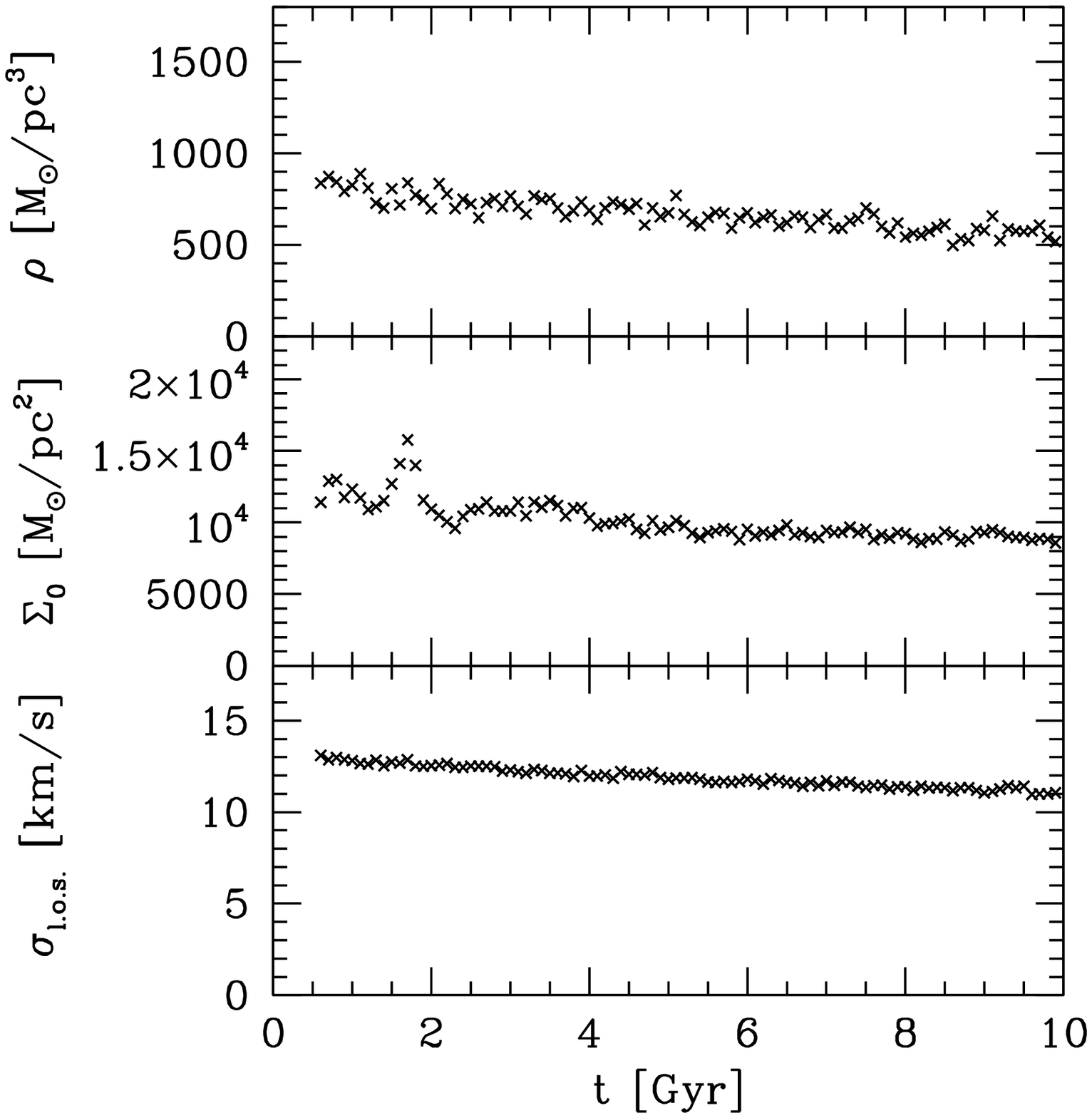}
    \caption{Time evolution of the central values of space
      density, surface density and line-of-sight velocity
      dispersion of merger object E01.} 
    \label{fig:cv}
  \end{center}
\end{figure}

The mass of the merger object is $M_{\rm mo} = 1.12 \cdot
10^{7}$~M$_{\odot}$ which corresponds to 91.5\% of the total
mass of the super-cluster.  The remaining mass is lost due to
internal heating from cluster--cluster collisions and subsequent
loss due to the tide.  The mass of the merger object decreases
with every perigalacticon passage.  The averaged mass-loss rate is
$5.7 \cdot 10^{5}$~M$_{\odot}$/Gyr.  After 10~Gyr the total mass
of the remaining object is $0.59 \cdot10^{7}$~M$_{\odot}$.

By $t=1$~Gyr, the gravitationally bound particles comprising the
merger object have a diameter of several~100~pc with an elliptical
envelope ($\varepsilon \approx 0.6$) but a spherical and very dense
core (Fig.~\ref{fig:c600}).  Bound particles have negative binding energies 
relative to the merger object.  The half-mass radius $r_{\rm h}=
54.6$~pc and $r(90\%)=149.2$~pc holds 90~\% of the mass
(Fig.~\ref{fig:lag}).  Examining the surface density profile
(Fig.~\ref{fig:surf}), the core radius $r_{\rm core}=10$~pc (the
radius where the surface density drops to half the central value).
The tidal radius at perigalacticon, estimated from Eq.~\ref{eq:tidal}
is $r_{\rm tidal} \approx 188$~pc.  The morphology of the object stays
almost constant during the whole computation time.  At $t=10$~Gyr the
half-mass radius is $r_h=r(50\%) = 41.9$~pc, $r(90\%) = 146.3$~pc,
$r_{\rm core} = 9.5$~pc and $r_{\rm tidal} = 152$~pc, as shown in
Fig.~\ref{fig:lag}. 

The structure of the merger object (MO) can be best fitted by a
double exponential profile.  The MO shows a very dense core
surrounded by an exponential envelope.  Defining the profile as
\begin{eqnarray}
  \label{eq:rho}
  \rho(r) & = & \rho_{0} \, \exp \left( - \frac{r}
    {r_{\rm exp}} \right),
\end{eqnarray}
the central density at $t=1$~Gyr is $\varrho_{0} = 1030 \pm
40$~M$_{\odot}$/pc$^{3}$ and the exponential scale length of the core
is $r_{\rm exp} = 6.7 \pm 0.2$~pc.  The envelope has $\varrho_{0} =
15.3 \pm 0.7$~M$_{\odot}$/pc$^{3}$ and $r_{\rm exp} = 26.0 \pm
0.3$~pc.  After 10~Gyr the values change to $\varrho_{0} = 750 \pm
15$~M$_{\odot}$/pc$^{3}$ and $r_{\rm exp} = 7.1 \pm 0.1$~pc for the
core and $\varrho_{0} = 6.6 \pm 0.5$~M$_{\odot}$/pc$^{3}$ and $r_{\rm
exp} = 27.8 \pm 0.8$~pc for the envelope.  Fig.~\ref{fig:dens} shows
the data points for $t =$ 1, 5 and 10~Gyr and the fitted functions for
$t=$ 1 and 10~Gyr.  Plotted as vertical lines are the approximate
tidal radii (Eq.~\ref{eq:tidal}).  The tidal extension of the MO
beyond the tidal radius is clearly evident.  

The surface density  is well fitted by a single de-Vaucouleur
profile,
\begin{eqnarray}
  \label{eq:vauc}
  \Sigma(r) & = & \Sigma_{\rm v} \cdot \exp \left( - 7.67 \left[
      \left( \frac{r}{r_{\rm v}} \right)^{1/4} -1 \right] \right),
\end{eqnarray} 
with $\Sigma_{\rm v} = 1500$~M$_{\odot}$/pc$^{2}$ and $r_{\rm v} = 18$~pc 
(fitted at $t=1$~Gyr) (Fig.~\ref{fig:surf}).  
Fitting the profile at $t=10$ Gyr with two
exponentials, results in a central surface density of $10600 \pm
300$~M$_{\odot}$/pc$^{2}$ and exponential scale lengths of $10.8 \pm
0.4$ and $35.7 \pm 0.5$~pc, respectively. 

\begin{figure*}
  \begin{center}
    \epsfxsize=15cm
    \epsfysize=11.5cm
    \epsffile{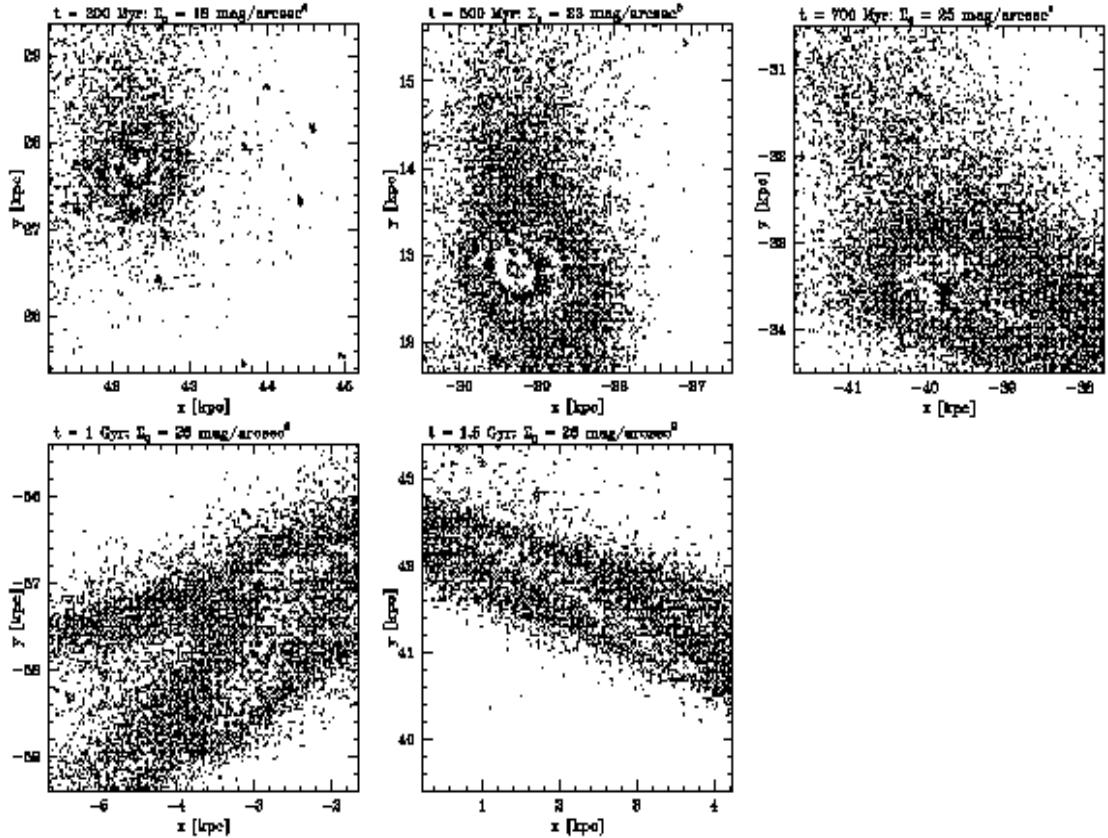}
    \caption{Evolution and destruction of the low-mass merger
      object R06.  Shown are contour-plots with 20~pc
      resolution.  Snapshots are at $t =$ 200, 500, 700, 1000 and
      1500 Myr.  Highest surface-brightness (adopting $M/L=1$)
      of the MO is indicated above each panel.} 
    \label{fig:r6c}
  \end{center}
\end{figure*}

Measuring the line-of-sight velocity dispersion shows
that the central velocity dispersion is about 13~km/s at the
beginning and drops down to about 11~km/s after 10~Gyr.  The
velocity dispersion profile can be fitted by two exponentials 
(Fig.~\ref{fig:losv}).

The central density, surface density and velocity dispersion decrease
only very slowly and linearly with time (Fig.~\ref{fig:cv}).  The
merger object is thus massive and stable enough to survive for a
Hubble time despite the very large orbital eccentricity.


\subsection{Low-concentration super-cluster on an eccentric
  orbit} 
\label{sec:eccld}

If the initial super-cluster has a sufficiently small $\alpha$ and is
subject to a strong tidal field, the merging efficiency can be low.
The resulting merger object has a smaller mass and is less
concentrated than model~E01 above.  Merger objects like this are
easily disrupted at their first perigalacticon passage forming an
unbound stream of stars and surviving star clusters moving along
almost identical orbits.  However, a density enhancement survives for
a substantial time, being composed of a core of particles with similar
phase-space coordinates that may perhaps be identified as a dSph
satellite (Kroupa 1997; Klessen \& Kroupa 1998).

In a calculation of a model with a Plummer radius of the
super-cluster of 300~pc orbiting on an eccentric orbit with
$e=0.33$ only a small merger object R06, made out of 4 star
clusters, forms.  It has a mass of about $0.35 \cdot
10^{7}$~M$_{\odot}$ and gets dissolved after the first
perigalacticon passage (Fig.~\ref{fig:r6c}).  The `central'
density drops from 100~M$_{\odot}$/pc$^{3}$ down to
0.01~M$_{\odot}$/pc$^{3}$.  The stars of this density enhancement
cover an area a few kpc in diameter.  Despite the fact
that the stars do not form a bound entity, the density enhancement
is still detectable after a few Gyrs.


\subsection{Evolution in the Kormendy diagram}
\label{sec:evol}

\begin{figure*}
  \begin{center} 
    \epsfxsize=15cm 
    \epsfysize=15cm
    \epsffile{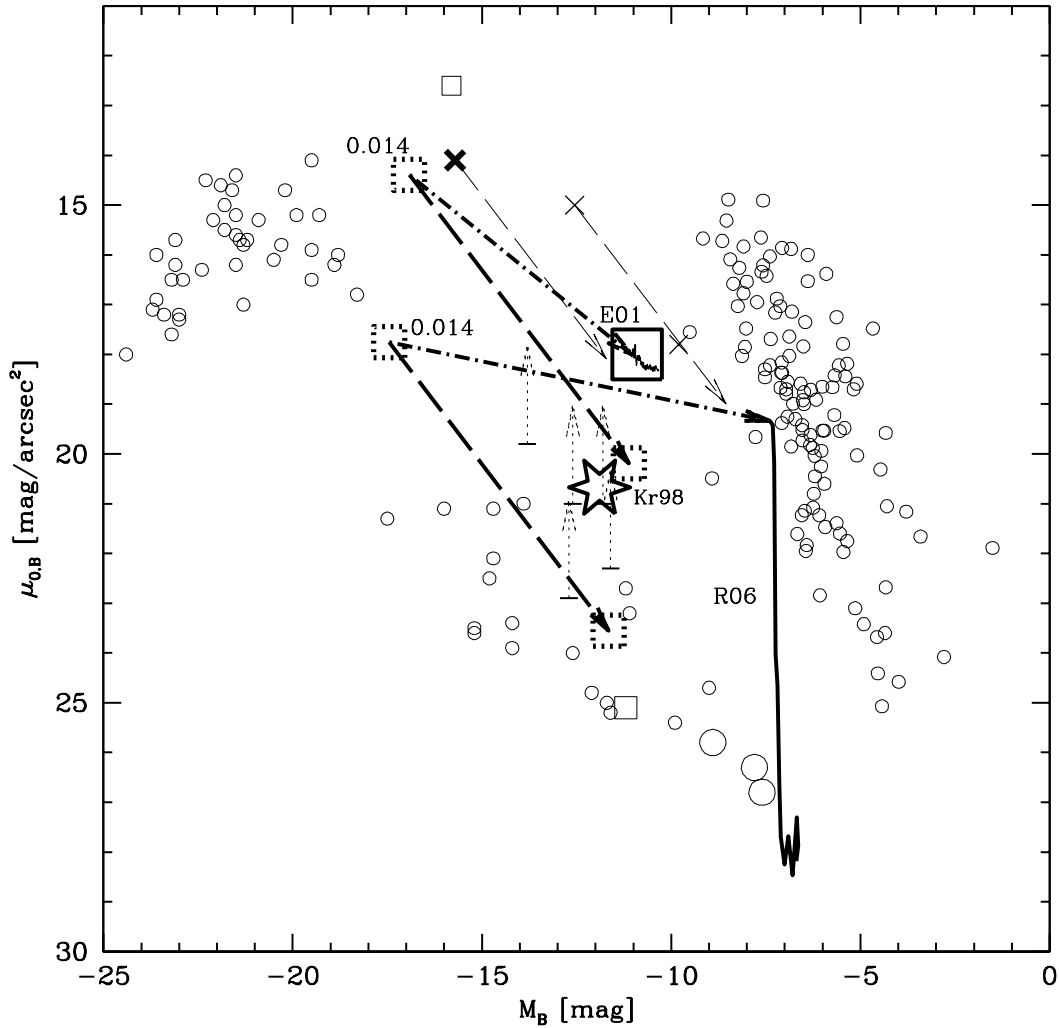} 
    \caption{Known stellar systems are shown as faint open
      symbols, apart from highlighted knot~S
      (Fig.~\ref{fig:k_obs}).  Models~E01 and~R06 are shown at an
      age of about 5~Myr as the thick dotted squares assuming
    $M/L_{\rm B}=0.014$.  Fading
      without morphological evolution to $M/L_{\rm B}=3$ would
      place these at the tips of the thick long-dashed arrows.
      The actual dynamical evolution is plotted as solid curves assuming
      the merger objects have $M/L_{\rm B}=3$.} 
    \label{fig:k_theor} 
  \end{center} 
\end{figure*}

The models discussed in Sections~\ref{sec:circ}-\ref{sec:eccld}
are placed in the Kormendy diagram for a comparison with known
stellar systems (Fig.~\ref{fig:k_theor}).

The models on circular orbits (Section~\ref{sec:circ}) overlap
the Kr98 prediction shown as the six-pointed open star assuming
$M/L_{\rm B}=3$.  These models do not evolve noticeably over a
Hubble time, apart from slight fading due to stellar
evolution. 

The evolution of the dense super-cluster on a highly eccentric
orbit (Section~\ref{sec:ecchd}) is shown for ages between 600~Myr
and 10~Gyr as the solid curve in the thick box labelled
E01.  Again, the evolution due to tidal harassment is not
significant.  The position of the equivalent 5~Myr old stellar
cluster with $R_{\rm Pl}=50$~pc and $M=1.23\times10^7\,M_\odot$
is shown as the upper thick dotted open square labelled with the
mass-to-light ratio $M/L_{\rm B}=0.0144$ (fig.~7 in Smith \&
Gallagher 2001) and assuming $B-V=0$.  Fading without
morphological evolution (Eq.~\ref{eqn:morph}) to $M/L_{\rm B}=3$
places such a cluster at the end of the thick long-dashed arrow
in the vicinity of the Kr98 prediction, while the actual
evolution (photometric and dynamical) places it in the open box
which lies in the vicinity of the fading vector of knot~S
(thick cross).

The evolution of the low-density super-cluster on an eccentric orbit
(Section~\ref{sec:eccld}) is shown for ages spanning from 200~Myr to
1.5~Gyr as the thick solid line labelled R06. By 1.5~Gyr a remnant
object stabilises in the vicinity of the known dSph satellites, and
sports a population of~16 co-streaming globular clusters.  An
equivalent stellar system that has the same Plummer density profile
and mass as the initial super-cluster R06 appears at an age of about
5~Myr at the position of the lower thick dotted box labelled as
$M/L_{\rm B}=0.014$.  It would resemble a low-mass dIrr, since the
distribution of brilliant young star clusters would appear patchy in
such a system. Hunter, Hunsberger \& Roye (2000) discuss observable
signatures of tidal-dwarf galaxies which may be related to this class
of object.  Fading to $M/L_{\rm B}=3$ without morphological evolution
would place this object at the end of the thick long-dashed arrow, in
the vicinity of bright dSph satellites.

Finally, from Fig.~\ref{fig:k_obs} we note that knot~430 (age
$\approx10$~Myr), which is less massive than~S but has a similar
central surface density, may evolve via fading slightly past knot~225
(age $\approx500$~Myr) to $M/L_{\rm B}=3$ if it undergoes no
morphological evolution (eqn~\ref{eqn:morph}).  Since the Milky-Way
globular cluster $\omega$~Cen also lies in the vicinity of the final
position of faded knot~430, we may assume that the precursor of
$\omega$~Cen may have looked similar to knot~430, or an object even
more massive lying closer to knot~S if dynamical evolution is taken
account of (Fig.~\ref{fig:k_theor}).  $\omega$~Cen may thus have been
born as a super-cluster containing hundreds of star clusters.  Such a
scenario may also be consistent with the unusual metallicity
distribution in $\omega$~Cen (e.g.\ Smith et al.\ 2000), which is
being taken as evidence that this unusual cluster may in fact be a
compact dwarf galaxy. Such an object encompassing about one hundred~pc
most probably forms over 5-10~Myr (Elmegreen et al.\ 2000) so that
self-enrichment through supernovae is very likely.  This scenario
merits further study.


\section{Conclusions}
\label{sec:conclus}

When gas-rich galaxies interact strongly the supersonically
colliding gas clouds are pressurised to such an extend that they
profusely form stars in super-clusters that have masses ranging
from $10^7$ to $10^8\,M_\odot$ and dimensions of a few hundred~pc.
Such sub-structures are observed to form in tidal arms in
computer models of colliding galaxies (Barnes \& Hernquist
1992; Elmegreen, Kaufman \& Thomasson 1993), and are also evident
in high-resolution HST images of the Antennae galaxies (Whitmore
et al. 1999).

These reveal the super-clusters to be composed of many star-clusters,
thus essentially conforming to the hierarchical star-formation picture
discussed by Elmegreen et al. (2000), according to which sub-structure
exists on every scale.  This is evident in local embedded clusters
(Megeath et al. 1996; Kaas \& Bontemps 2001) as well as the region
with a radius of a few hundred~pc around 30~Dor in the Large
Magellanic Cloud, which contains many star clusters smaller and less
massive than the central cluster R136.

It is thus to be expected that globular clusters will also have
formed heavily sub-structured, i.e.  essentially composed of many
smaller clusters.  Extreme cases may lead to $\omega$~Cen, which
can be interpreted to form the borderline between massive star
clusters and small dwarf galaxies.  The present findings
(Section~\ref{sec:evol}) lend credence to this notion, which may
also allow an explanation of the metallicity spread in
$\omega$~Cen.

Even more massive and larger super-clusters, such as are observed for
example in the Antennae galaxies (knot~S), evolve through a phase of
violent star-cluster interactions to stellar systems with relaxation
times longer than a Hubble time and so form compact dwarf galaxies
that are stable even when subject to strong tides.  Examples of such
systems may have been discovered recently by Phillipps et al. (2001)
in the Fornax galaxy cluster as ultra-compact dwarf galaxies. \\


\noindent {\bf Acknowledgements:}\\

\noindent
MF acknowledges financial support through DFG-grant FE564/1-1.  Part
of this project (parallelisation of the code and use of
high-performance computers) was carried out at the Edinburgh Parallel
Computing Centre (EPCC) through the TRACS-programme.  The
TRACS-program\-me is a scheme of the European Community: Access to
Research Infrastructure action of the Improving Human Potential
Programme (contract No.\ HPRI-1999-CT-00026) which allows young PhD-
or post-doctoral scientists to learn parallel computing by offering
computing time and support for their projects.  MF thanks the staff of
EPCC for their support.



\end{document}